\documentclass{pasa}
\usepackage[usenames,dvipsnames]{xcolor}

\usepackage{aas_macros}
\usepackage{amssymb,amsmath}
\usepackage{bm}
\usepackage{graphicx}

\title[CFS instability and GW multipolar expansion]
{Radiation driven instability of rapidly rotating relativistic stars: criterion and evolution equations via multipolar expansion of gravitational waves}

\author[A. I. Chugunov]
{A.~I.~Chugunov\\
	\affil{
		Ioffe Institute, St Petersburg, Russia. }
}
\jid{pasa}

\doi{10.1017/pas.\the\year.xxx}
\jyear{\the\year}

\usepackage[authoryear]{natbib}

\usepackage{aas_macros}
\usepackage{hyperref}
\hypersetup{colorlinks,citecolor=blue,linkcolor=blue,urlcolor=blue}

\begin{document}


\begin{abstract}
I suggest a novel approach for deriving evolution equations for  rapidly rotating
relativistic stars affected by radiation-driven Chandrasekhar-Friedman-Schutz (CFS) instability.
This approach is based on the multipolar expansion of gravitational wave emission  and appeals to the global physical properties of the star (energy, angular momentum, and thermal state), but not to canonical energy and angular momentum, which is traditional. 
It leads to simple derivation of the CFS instability criterion for normal modes and the evolution equations for a star, affected by this instability. 
The approach also gives a precise form to simple explanation of the CFS instability: it occurs  when two conditions met: (a) gravitational wave emission removes angular momentum from the rotating star (thus releasing the rotation energy) and (b)  gravitational waves carry less energy, than the released amount of the rotation energy.
To illustrate the results I take the r-mode instability in slowly rotating Newtonian stellar models as an example. It leads to evolution equations, where the emission of gravitational waves directly affects the spin frequency, being in line with the arguments by \cite{lu01b}, but in contrast to  widely accepted equations by \cite*{lom98,hl00}. According to the latter, effective spin frequency decrease is coupled  with dissipation of
unstable mode, but not with the instability as it is. This 
problem, 
initially stressed by \cite{lu01b},  is shown to be superficial, and arises as a result of specific definition of the effective spin frequency, applied by \cite{lom98,hl00}.
Namely, it is shown, that if this definition is taken into account properly, the evolution equations coincide  with obtained here in the leading order in mode amplitude. 
I also argue that the next-to-leading order terms in evolution equations (which differ for \citealt{olcsva98} and \citealt{hl00}) require clarification and thus it would be more self-consistent to omit them.
\end{abstract}

	
\maketitle

\section{Introduction}

\cite{andersson98,fm98} demonstrate that all rotating stars are
unstable with respect to excitation of r-modes (similar to Earth's Rossby waves controlled by the Coriolis force)  at any
rotation rate, if  dissipation is neglected. It is a particular case of Chandrasekhar-Friedman-Schutz (CFS) instability
(\citealt{chandrasekhar70a,fs78a,fs78b}).  Evolution
equations for dissipative neutron star, affected by the r-mode
instability, were derived by \cite{olcsva98} (see also
\citealt{levin99,hl00,as14,gck14b}), assuming slow rotating Newtonian stellar model.
These equations are widely applied in literature and many observational consequences were predicted (see \citealt{haskell15,cgk17}, for recent reviews). In particular, the r-mode instability can limit spin frequencies of neutron stars (\citealt{bildsten98,aks99}), generate potentially observable gravitational waves (e.g.\ \citealt{owen10}), and lead to anti-glitches in millisecond pulsars and neutron star, accreting in low mass X-ray binaries (\citealt*{kgc16}) and even  formation of additional class of hot rapidly rotating neutron stars -- HOFNARS, -- which could reveal itself by stable thermal emission from surface, but do not accrete (\citealt*{cgk14}). 
Confronting observations with predictions of r-mode instability theory one can put important constraints on the physics of neutron stars, including properties of their depths    (see e.g., \citealt*{hdh12}; \citealt{haskell15,cgk17}).

To a great extent the above results are based  on the canonical energy formalism and Lagrangian perturbation theory formulated by  \cite{fs78a, fs78b} and generalized for relativistic case by \cite{Friedman78}.
This theory is, of course, mathematically strict, 
but it is  but rather complicated: 
\cite{fs78a, fs78b} reveal class of Lagrangian displacements, -- called trivials, -- which do not modify physical variables and introduce another class of Lagrangian displacements -- called of canonical, -- which are orthogonal to the trivials. They also introduce canonical energy
functional for Lagrangian perturbations and demonstrate that perturbations described by  canonical Lagrangian displacement with negative canonical energy are unstable with respect to gravitational radiation (in absence of viscosity). 
It is worth to note that canonical energy  can be equal to the physical change of energy under certain conditions (e.g., growth of normal mode under action of radiation reaction force in absence of viscosity), but generally it is not a case (\citealt{fs78a, fs78b}). 
This feature leads to conceptual critique (\citealt{lu01b}, see also footnote \ref{foot_NegMom}) of the first derivation of evolution equations for CFS unstable star (\citealt{olcsva98}). Objections  were rather convincingly  replied to by \cite{hl00}, who suggest slightly modified evolution equations.
Currently both versions 
are applied by different authors (e.g., compare  \citealt{haskell15} and  \citealt{ms13}), but which of them is more precise?

To see the problem in a different light, I  suggest another approach, which as I believe, is also useful from methodological point of view. Namely, I avoid  to use Lagrangian perturbation theory and canonical energy and angular momentum of perturbations,  but  deal  with global physical properties of the star (energy, angular momentum, and the thermal state)
and consider instability of normal modes.%
\footnote{ It is worth to note, that Lagrangian perturbation theory developed by  \cite{fs78a, fs78b,Friedman78}  allows to consider instability of the initial data with no assumptions concerning  the existence or completeness of normal modes. }
The key point is Eq.\ (\ref{rate_GW_general}), which  follows from the multipolar expansion of gravitational wave emission (\citealt{Thorne80}) and couples the rates of change of the energy and the angular momentum.
It  naturally reveals physics of CFS instability: emission of gravitational waves should remove angular momentum from the star, thus releasing  the rotation energy. The instability occurs if gravitational waves carry less energy than the released amount of the rotation energy (see first equality in Eq.\ \ref{Eex_GW}).

Similar explanation was given, e.g., by \cite{andersson98,ak01}, but that was rather heuristic arguments than strict proof (e.g., \citealt{fs14}).%
\footnote{Let me note that earlier  it was typically assumed that unstable perturbation (as it is) decreases physical angular momentum. However, as noted by \cite{lu01b}, it is rather misleading, and, for example,
such perturbations 
can exist in a star with same physical angular momentum as unperturbed configuration (see also thorough discussion of the `wave-momentum' myth by \citealt{mcintyre81}).
\label{foot_NegMom}}
Explicit reference to Eq.\ (\ref{rate_GW_general}) allows to formulate these arguments in a precise form and not only derive instability criterion, but also obtain general evolution equations for CFS unstable star (see Sec.\ \ref{Sec_EvolSimple}). It is worth noting, that this derivation does not require simplifying assumptions of slow rotation or Newtonian gravitation, but is valid in full general relativity framework.
Additional advantage of  derivation in Sec.\ \ref{Sec_EvolSimple}  is that it allows straightforward generalization for   superfluid   neutron stars, because it exploits the global properties of the star (energy and angular momentum), but not complicated structure of internal perturbations. 

To apply the evolution equations, derived in Sec.\ \ref{Sec_EvolSimple} one, of course,  should deal with detailed description of internal perturbations to calculate efficiency of the gravitational radiation and dissipation timescales. It is very complicated problem, which can be crucially affected by the neutron star core composition  (e.g., \citealt{Jones01_comment,lo02,no06,as14_msp}), superfluidity (\citealt{yl03a,ly03,agh09,hap09,gck14a,kg17}), crust-core coupling (\citealt{rieutord01,lu01,ga06a,ga06b}), in-medium effects (\citealt{kv15}), and, of course, general relativity (e.g., \citealt*{kojima98,laf01,yl02,yl03b,rk02,lfa03,law04,kgk10}).
However, I leave these problems beyond the scope of this paper, 
because they provide parameters for the evolution equations, but do not modify their form.

In section \ref{Sec_slow} I illustrate the general evolution equations on the example of slow rotating Newtonian stellar model. In particular, it is shown that the gravitational radiation directly leads that the star spins down, being thus in agreement with arguments by \cite{lu01b}.
In section \ref{Sec_Comp} these results are compared with  widely applied equations by \cite{olcsva98,hl00}. At the first glance my equations differ from those in both of papers: latter  attribute decrease of the effective spin frequency (as introduced by \citealt{olcsva98,hl00}) to the dissipation of the unstable mode, but not with instability as it is. I demonstrate that this difference is associated with definition of the effective frequency  and just as it is taken into account, the evolution equations agree  in the leading order in mode amplitude. I also argue that the next-to-leading order terms, which differ for \cite{olcsva98} and \cite{hl00}, was not yet derived accurately and should be omitted.
I conclude in section \ref{Sec_Sum}.

\section{CFS instability and evolution of unstable star}
\label{Sec_EvolSimple}

In this section I consider rotating
relativistic star in its asymptotic rest frame; their total
mass-energy $E$ and angular momentum $J$ are well defined
(see, e.g., Sec.\ 19 in the textbook by \citealt*{MTW}).
 For  given angular momentum $J$, uniform
rotation corresponds to the minimal energy (at fixed baryon
number), which is the rotational energy $E_\mathrm{rot}$ (e.g.\
\citealt{bl66,hs67,Stergioulas03}). Corresponding spin frequency is (it follows, e.g., from variational principle by \citealt{hs67})
\begin{equation}
    \Omega=\frac{\partial  E_\mathrm{rot}}{\partial J}.
    \label{Omega}
\end{equation}
If star is perturbed, but the total
angular momentum does not changed, the  energy $E$ exceeds
rotational energy ($E>E_\mathrm{rot}$), leading to positively defined excitation energy:
\begin{equation}
    E_\mathrm{ex}=E-E_\mathrm{rot}. \label{E_ex}
\end{equation}
I would like to stress, that $E_\mathrm{ex}$  should not be confused with non-positively-defined
canonical energy, introduced by \cite{fs78a,fs78b}.

As shown by \cite{Thorne80}, the rate of changes of the energy $\dot E^\mathrm{GR}$ and
angular momentum 
$\dot
J^\mathrm{GR}$ due to emission of gravitational waves 
can be expressed as sums over multipolar contributions,
which comes from  expansion of radiation field in the local wave zones. 
As it follows from Eqs.\ (4.16) and (4.23) by \cite{Thorne80}, for perturbations $\propto \mathrm e^{\imath (\omega
	t+m\phi)}$,
these rates are coupled by the equation%
\footnote{This
equation is well known for electromagnetic
waves, see e.g.\ Sec.\ 9.8 in textbook by \cite{JacksonCED}.}
\begin{equation}
    -\frac{\omega}{m} \dot J^\mathrm{GR}=\dot
    E^\mathrm{GR}. \label{rate_GW_general}
\end{equation}
Gravitational wave emission
removes energy from the system, thus $\dot E^\mathrm{GR}<0$.
Sign of rate of change of angular momentum  is
determined by $\omega/m$. The rate of change of the excitation energy
$E_\mathrm{ex}$ is
\begin{equation}
  \dot E^\mathrm{GR}_\mathrm{ex}=\dot E^\mathrm{GR}-\dot E^\mathrm{GR}_\mathrm{rot}
  =\left(1+\frac{m\Omega}{\omega}\right) \dot E^\mathrm{GR}.
\label{Eex_GW}
\end{equation}
Here $\dot E^\mathrm{GR}_\mathrm{rot}=\Omega \dot J^\mathrm{GR}$.
The excitation energy is increased by emission of
gravitational waves if and only if $(1+m\Omega/\omega)<0$. 
This condition is equal to well known  criterion
of CFS instability (see e.g. \citealt{fs78a,Friedman78,ak01,fs14}): the prograde mode pattern in the inertial
frame ($-\omega/m>0$), but retrograde mode pattern in the frame, corotating
with the star ($-\Omega-\omega/m<0$). Thus the above discussion proves the CFS instability criterion for normal modes without appeal to the Lagrangian perturbation theory. 

To describe the evolution of CFS unstable star, I parametrize it state by three parameters: (i) total angular momentum $J$, (ii) the mode energy $E_\mathrm{ex}$, and (iii) thermal state. The latter can be characterized by 
temperature in the stellar centre $T$, because neutron stars are almost isothermal (e.g., \citealt*{plps04,gkyg05}) because of high thermal conductivity in their depths (see, e.g., \citealt*{sbh13} for recent results).

Evolution of angular momentum due to emission of gravitational waves 
is described by Eq.\ (\ref{Eex_GW}) 
\begin{equation}
  \dot J^\mathrm{GR}
= -\frac{m}{\omega+m\Omega} \dot
  E^\mathrm{GR}_\mathrm{ex}. \label{J_evol_gen}
\end{equation}
In this
equation, which is applicable for
any oscillation mode
(in particular for r-modes)
at any spin frequency and even for general relativistic
(not Newtonian) stellar models, $\Omega=\Omega(J)$ is determined by Eq.\
(\ref{Omega}). 

Evolution of the mode energy is associated with energy pumping by gravitational waves $\dot E^\mathrm{GR}_\mathrm{ex}$ and energy losses due to dissipation $\dot E^\mathrm{dis}_\mathrm{ex}$
\begin{equation}
\dot E_\mathrm{ex}=\dot E^\mathrm{GR}_\mathrm{ex}+\dot
     E^\mathrm{dis}_\mathrm{ex}.
     \label{Eex_evol_gen}
\end{equation}

Finally, the thermal evolution of star follows
\begin{equation}
  C \dot T=
  - \dot E^\mathrm{dis}_\mathrm{ex}
   -L_\mathrm{cool}.
 \label{thermal_gen}
\end{equation}
Here 
$L_\mathrm{cool}$, and $C$ are 
total cooling power (neutrino and thermal emission from
surface) and heat capacity of the star
respectively. One can also add torques, which are not associated with r-modes (e.g., accretion spin up, e.g.,\ \citealt{gl79b,wang95} or magnetic braking, e.g.,\ \citealt*{bgi93}) to the angular momentum evolution equation (\ref{J_evol_gen}) and additional heating [e.g., accretion-induced deep crustal heating (\citealt*{bbr98}) or internal heating in millisecond pulsars (\citealt*{Alpar_etal84,Reisenegger95,gkr15})] to the thermal evolution equation (\ref{thermal_gen}).

 To apply equations (\ref{J_evol_gen})--(\ref{thermal_gen}) one should specify properties of the mode:
$\omega$, $\dot E^\mathrm{GR}_\mathrm{ex}$
and
 $\dot E^\mathrm{dis}_\mathrm{ex}$
as function of $E_\mathrm{ex}$ and $J$, which is, of course, very complicated problem, especially for relativistic stellar models (see e.g.\ \citealt{lfa03,Kastaun11}). However, for slow rotating Newtonian stellar models these parameters can be easily extracted from literature (yet, depending on the microphysical assumptions, see e.g., \citealt{haskell15}) and  in the next section they are applied to illustrate equations (\ref{J_evol_gen})--(\ref{thermal_gen}). %

\section{Evolution of r-mode unstable neutron star within slow rotating Newtonian stellar models}
\label{Sec_slow}

Here I restrict  myself to slow rotating Newtonian stellar models,
which are commonly used  to study CFS
instability in neutron stars (see, e.g., \citealt{haskell15} for
recent review). In this case, Eq.\ (\ref{Omega}) can be written as $\Omega=J/I$, where moment of inertia $I$ does not depend on $J$. The most unstable mode is
r-mode with $l=m=2$ (e.g., \citealt{lom98}), with the
frequency 
\begin{equation}
 \omega=-\frac{(m-1)(m+2)}{m+1}\Omega=-\frac{4}{3}\Omega.
 \label{om_r}
\end{equation}
The  first order Eulerian perturbations of
velocity can be written as follows (e.g., \citealt*{pbr81}):
\begin{equation}
 \delta^{(1)} \bm v=
 \alpha R\Omega \left(\frac{r}{R}\right)^m \bm Y_{mm}^B
 \exp^{i\omega t}
       \label{delta_v}
\end{equation}
Here  $\alpha$ is dimensionless mode amplitude and 
\begin{equation}
\bm Y_{lm}^B=\frac{1}{l(l+1)} r
  \nabla \times (r\nabla Y_{lm})
\end{equation}
is magnetic-type vector spherical harmonic (see e.g., \citealt*{vms88}).
The excitation energy for r-mode can
be written in the form
\begin{eqnarray}
    E_\mathrm{ex}&=&\int \frac{\rho \delta  v^2}{2} \mathrm d ^3
    \bm r
    =\int \frac{\rho [\delta^{(1)} \bm v]^2}{2} \mathrm d ^3
    \bm r
    \nonumber \\
    &+& \int \rho  \bm v_0 \delta^{(2)} \bm v  \mathrm d ^3
    \bm r
    +\mathcal O(\alpha^3), \label{E_slow_gen}
\end{eqnarray}
Here $\delta v^2=(\bm v)^2-(\bm v_0)^2=2 \bm v_0\delta\bm v+(\delta\bm v)^2$ is perturbation of squared  velocity and  $\delta \bm v=\sum_i\delta ^{(i)} \bm v$ is total
perturbation of velocity, presented as a sum over orders in
$\alpha$ [i.e., $\delta ^{(i)} \bm v=\mathcal O(\alpha^i)$]. The integral is taken over stellar volume. I neglect also density pertrubations, because they are of the second order in $\Omega$ (e.g., \citealt{lom98}).
The second term in Eq.\
(\ref{E_slow_gen}) depend on the
second order velocity perturbation $\delta^{(2)}\bm v$, but because of the finite velocity at the unperturbed state $\bm v_0$, it contributes to the energy at the same order as first order perturbations (\citealt{fs78a}).%
\footnote{The linear contribution from the first order perturbation vanished after integration over stellar volume, e.g., \cite{lu01b}.} However, only axysimetric part   $\delta^{(2)}_\mathrm{sym}\bm v$ can contribute to the integral,%
\footnote{For given first order solution, the asymmetric part $\delta^{(2)}_\mathrm{sym}\bm v$ is 
determined up to arbitrary cylindrically stratified differential rotation
(e.g.,\ \citealt{Sa04}).}
but the definition of excitation energy (Eq.\ \ref{E_ex}) supposes that
the perturbed state has the same angular momentum as
unperturbed star, constraining $\delta \bm v$:
\begin{equation}
    \delta J=\int \rho \left[\delta \bm v\times \bm
    r\right]\mathrm d^3 \bm r=0. \label{J_constraint}
\end{equation}
As far as unperturbed state is uniform rotation
$\bm v_0=\bm \Omega\times \bm r$, the contribution of
$\delta^{(2)}_\mathrm{sym} \bm v$ to the energy should vanish in Eq.\ (\ref{E_slow_gen})  
at the second order in $\alpha$. Thus, the second order excitation energy
is determined exclusively by the first order perturbations and equals to
the kinetic energy in the system corotating with star. It  can be written as follows (see,
e.g., \citealt{lom98}):
\begin{equation}
    E_\mathrm{ex}=\frac{1}{2} \alpha^2 \Omega^2 R^{-2m+2} \int _0^R \rho r^{2m+2}\mathrm d^3
    \bm r
    \label{E_ex_alpha}
\end{equation}

The instability timescale
\begin{equation}
	\tau^\mathrm{GR}=-2 \frac{E_\mathrm{ex}}{\dot E^\mathrm{GR}_\mathrm{ex}} \label{tau_GR_gen}
\end{equation}
can be calculated via multipolar expansion of gravitational
radiation for Newtonian sources (see Sec.\ V.C in
\citealt{Thorne80}), as it was done by \cite{lom98}:
\begin{eqnarray}
\frac{1}{\tau^\mathrm{GR}}
&=&-\frac{32\pi G\Omega^{2m+2}}{c^{2m+3}}
\,\frac{(m-1)^{2m}}{[(2m+1)!!]^2}
\,\left(\frac{m+2}{m+1}\right)^{2l+2}
\nonumber \\
&\times& \int_0^R \rho r^{2m+2} \mathrm d r.
\label{tau_GW}
\end{eqnarray}
This result agrees with analytic
treatment of r-mode instability up to second order in
oscillation amplitude by \cite{fll16}. 

The dissipation rate
\begin{equation}
 \tau^\mathrm{dis}=-2\frac{E_\mathrm{ex}}{\dot E^\mathrm{dis}_\mathrm{ex}} \label{tau_dis_gen}
\end{equation}
should be specified for each certain model of dissipation
(shear viscosity, mutual friction, etc.). Note, 
internal dissipative processes can not affect  
total angular momentum of the star, thus the rotational energy is conserved and dissipation time scale can be estimated from dissipation rate of the total energy $\dot E^\mathrm{dis}$ (i.e.\ one can substitute $\dot E^\mathrm{dis}$ instead of $\dot E^\mathrm{dis}_\mathrm{ex}$ in Eq.\ \ref{tau_dis_gen}). For example, the contribution of the shear viscosity $\eta$ to the dissipation rate is (\citealt{lom98}):
\begin{equation}
\tau^\mathrm{S}=(m-1)(2m+1)\frac{\int_0^R \eta r^{2l} \mathrm d r}
{\int_0^R \rho r^{2l+2} \mathrm d r}.
\end{equation}
Introduction of
these timescales gives $\dot E^\mathrm{GR}_\mathrm{ex}$,
and $\dot E^\mathrm{dis}_\mathrm{ex}$ as functions of $J$ and $E_\mathrm{ex}$, allowing thus to rewrite Eqs.\ (\ref{J_evol_gen})
-- (\ref{thermal_gen}) in the form:
\begin{eqnarray}
  \dot \Omega &=&\frac{2\tilde Q \alpha^2}{ \tau^\mathrm{GR}(\Omega)}\Omega \label{dot_Om} \\
  \dot \alpha &=& -\left(\frac{1}{\tau^\mathrm{GR}}+\frac{1}{\tau^\mathrm{dis}}\right) \alpha \label{dot_alpha} \\
  C\dot T &=& \frac{\tilde J M R^2 \Omega}{\tau^\mathrm{dis}} \alpha^2
  -L_\mathrm{cool} \label{dot_T}
\end{eqnarray}
There we, following \cite{olcsva98}, introduce
dimensionless parameters
\begin{eqnarray}
 \tilde J&=&\frac{1}{MR^{2m}}\int_0^R \rho r^{2m+2} \mathrm d r\approx 1.64\times 10^{-2}, \label{tild_J} \\
 \tilde I&=&\frac{I}{MR^2}=\frac{8\pi}{3MR^2}\int_0^R \rho r^4 \mathrm d r\approx 0.261,  \label{tild_I}  \\
 \tilde Q&=&\frac{m(m+1)\tilde J}{4\tilde I}\approx 9.4\times 10^{-2}.  \label{tild_Q} 
\end{eqnarray}
The numerical values are for $m=2$ r-mode and Newtonian stellar model with polytropical EOS $P\propto \rho^{1+1/n}$ with $n=1$.
Note, in agreement  with  arguments by \cite{lu01b}, the spin down rate given by Eq. (\ref{dot_Om}) is directly associated with emission of gravitational waves.

The enhancement of the mode amplitude can be limited by nonlinear saturation (e.g., \citealt{btw07,bw13}; \citealt*{hga14}), which can be described by substitution of the effective dissipation rate $\tau^\mathrm{dis}=|\tau^\mathrm{GR}|$ instead of $\tau^\mathrm{dis}$ into all equations (\ref{dot_Om})--(\ref{dot_T}).

\section{Comparison with previous works}  \label{Sec_Comp}

At first glance, the evolution equations (\ref{dot_Om}-\ref{dot_T}) differ from equations derived by \cite{olcsva98,hl00} applied in vast majority of the papers dealing with the evolution of r-mode unstable NSs. Namely, in the leading order in mode amplitude the equations can be written in the form:%
\footnote{Here I neglect external torques and heating for the sake of simplicity.}
\begin{eqnarray}
\dot {\hat \Omega} &=&\frac{2\tilde Q \alpha^2}{ \tau^\mathrm{dis}}\hat\Omega \label{dot_Om_stand} \\
\dot \alpha &=& -\left(\frac{1}{\tau^\mathrm{GR}}+\frac{1}{\tau^\mathrm{dis}}\right) \alpha \label{dot_alpha_stand} \\
C\dot T &=& \frac{\tilde J M R^2\hat \Omega}{\tau^\mathrm{dis}} \alpha^2
-L_\mathrm{cool}. \label{dot_T_stand}
\end{eqnarray}
Here $\hat \Omega$ is the effective spin frequency, as introduced by \cite{olcsva98,hl00}, which  differs from $\Omega$, given by Eq.\ (\ref{Omega}).
Note, the rate of change of  $\hat \Omega$ is associated with dissipation timescale, but not with instability timescale as in Eq.\ (\ref{dot_Om}).

The reason of this difference originates from uncertainties in definition of mean spin frequency in CFS unstable neutron star arising from differential rotation, which can be generated in the star as a result of CFS instability (see, e.g., \citealt{Spruit99_DifRot,Rezzolla_etal00,lu01b,fll16}). 
Namely, \cite{olcsva98,hl00} assume that physical angular momentum associated with r-mode is equal to the canonical angular momentum
\begin{equation}
J_\mathrm{c}=-(3/2)\hat\Omega\tilde J MR^2 \alpha^2,
\end{equation}
and write total angular momentum as%
\begin{equation}
J=I\hat \Omega+J_\mathrm{c}. \label{J_olcsva}
\end{equation}
As a result, the effective spin frequency is
\begin{equation}
\hat \Omega= (1+\tilde Q\alpha^2)\Omega. \label{Om_hat}
\end{equation}
After this change of variables our equations agree with
results by \cite{olcsva98,hl00} in the leading order in
$\alpha$.

Equations derived by \cite{olcsva98,hl00} also contain terms, which are of the next-order in $\alpha^2$.
However I suppose that these terms (which do not agree for \citealt{olcsva98} and \citealt{hl00}) include only part of the required next-order corrections. 
To be brief, their equations are based on the linear (leading order) perturbation theory and hence cannot predict next-to-leading order corrections accurately. 
For example, both \cite{olcsva98} and \cite{hl00}  neglect corrections to the r-mode frequency
associated with differential rotation (assuming r-mode frequency to be $\omega=4\hat{\Omega}/3$ in all orders in $\alpha$).
However, as discussed  by \cite*{csy14}, the differential rotation $\delta \Omega$ can affect the spin frequency at order of $\mathcal{O}(\delta\Omega/\Omega)$.
As a result, the second order differential rotation, associated with excitation of r-modes (\citealt{Spruit99_DifRot,Rezzolla_etal00,lu01b,fll16}) 
leads to corrections to the r-mode frequency $\mathcal{O}(\alpha^2)$,
and consequently to corrections of the same order for the energy pumping rate, given by Eq.\ (\ref{J_evol_gen}).%
\footnote{In particular, to the best of my knowledge, there are no proofs that $\omega=4\hat \Omega/3$ is a better estimate for the $m=2$ r-mode frequency than $\omega=4 \Omega/3$, but these estimates differ at $\mathcal{O}(\alpha^2)$, see Eq.\ (\ref{Om_hat}). The similar argument holds true for $\tau^\mathrm{GR}$ and $\tau^\mathrm{dis}$, which, according to \cite{olcsva98,hl00}, should be calculated assuming uniform rotation frequency to be equal to $\hat \Omega$ (but not $\Omega$) in all orders in $\alpha$.}
Note, the numerical value of these corrections depend on the differential rotation profile (\citealt*{csy14}), which can be dramatically modified by the magnetic field in the star (see \citealt{Chugunov15,flrc17}, who also demonstrate that magnetic field windup by  differential rotation do not suppress r-mode instability thanks to back reaction of magnetic field) or viscosity.
To be accurate with the next order effects in the evolution equations, one needs additional modeling of the differential rotation profile.  Fortunately, these effects should be negligible in any case, because current models of nonlinear saturation (e.g., \citealt{btw07,bw13}; \citealt*{hga14}) predict low saturation amplitude $\alpha\lesssim 10^{-3}$ and actual observations constrain the r-mode amplitude in many potentially unstable neutrons stars even stronger (e.g.,  \citealt{ms13,as15_oscMSP,Schwenzer_etal_Xray,cgk17}). Thus I suggest to use leading order equations (\ref{dot_Om})--(\ref{dot_T}) (or equivalent equations \ref{dot_Om_stand}--\ref{dot_T_stand}) to model the evolution of r-mode unstable neutron stars and omit next-order terms.

\section{Summary and conclusions} \label{Sec_Sum}

Using multipolar expansion of gravitational waves, formulated by \cite{Thorne80},
I derive criterion of CFS instability for normal modes and evolution equations (\ref{J_evol_gen})--(\ref{thermal_gen}) for a star, affected by 
this instability.
The derivation does not
appeal to the canonical energy formalism by \cite{fs78a,fs78b}. 
Eqs.\ (\ref{J_evol_gen})--(\ref{thermal_gen}) describe evolution of angular momentum, mode energy and temperature and can be applied for relativistic star. They are illustrated by r-mode instability in slowly rotating Newtonian stellar model (Eqs.\ \ref{dot_Om}- \ref{dot_T}). 
In a latter case the evolution equations 
were earlier derived by \cite{olcsva98,hl00}. At first glance they equations differ from those derived here. However, it is shown that it is spurious: in the leading order in mode amplitude the only difference is definition of the (effective) spin frequency, entering into the equations ($\Omega$ or $\hat\Omega$, see Eq.\ \ref{Om_hat}).
Formulas, suggested by \cite{olcsva98,hl00} contain terms of the next order in the mode amplitude. However I argue that they do not include all of the required terms (as a result, these terms do not agree for \citealt{olcsva98} and \citealt{hl00}) and thus it would be more self-consistent to restrict consideration to the leading order evolution equations (Eqs.\ \ref{dot_Om}- \ref{dot_T}).

The difference in definition of the spin frequency ($\Omega$ and $\hat\Omega$, see Eq.\ \ref{Om_hat}) is associated with differential rotation, which can be 
generated 
during excitation of r-modes by CFS instability (e.g., \citealt{Spruit99_DifRot,hl00, Rezzolla_etal00,lu01b,fll16}). 
As far as observed frequency is associated with specific point at the surface of neutron star (e.g. magnetic pole), the observed spin frequency $\Omega_\mathrm{obs}$ can differ from both frequencies $\Omega$ and $\hat \Omega$.
Important advantage of the formalism, suggested in this work, that it can be naturally extended to include effects of differential rotation (because it explicitly deals with second order axisymmetric velocity perturbations, see Sec.\ \ref{Sec_slow}), and I plan to analyze them at the subsequent paper. 
As far as the spin down rate, given by Eq. (\ref{dot_Om}), qualitatively agrees with  arguments by \cite{lu01b}:
radiation of gravitation waves directly leads to spin down of the star (but not via dissipation of mode energy as it follows from \citealt{olcsva98,hl00}, see Eq.\ \ref{dot_Om_stand}), I believe that  $\Omega_\mathrm{obs}$  would be closer to $\Omega$, than to $\hat \Omega$. However, this conclusion should be checked by accurate calculations

\section*{Acknowledgements}
I'm  grateful to Misha Gusakov and Elena Kantor for valuable comments and discussions. This study was supported by the
Russian Science Foundation (Grant No. 14-12-00316).

\end{document}